# Two-dimensional atomic layer-molecule hybrid superconductors with controllable exchange coupling

Shunsuke Yoshizawa[1], Emi Minamitani[2], Saranyan Vijayaraghavan[1†], Puneet Mishra[1], Yasumasa Takagi[3], Toshihiko Yokoyama[3], Satoshi Watanabe[2], Tomonobu Nakayama[1] and Takashi Uchihashi[1*]

[1]International Centre for Materials Nanoarchitectonics (MANA), National Institute for Materials Science, 1-1, Namiki, Tsukuba, Ibaraki 305-0044, Japan

[2] Department of Materials Engineering, the University of Tokyo, 7-3-1, Hongo, Bunkyo-ku, Tokyo 113-8656, Japan

[3]Institute for Molecular Science, Myoudaiji Campus, 38 Nishigo-Naka, Myodaiji, Okazaki, Aichi 444-8585, Japan

[†]Present Address: CSIR-Central Electrochemical Research Institute, Karaikudi-630003, Tamil Nadu, India

*email: UCHIHASHI.Takashi@nims.go.jp

**The coexistence and competition of superconductivity and magnetism can lead to a variety of rich physics and technological applications. Recent discovery of atomic-layer superconductors and self-assembly of magnetic molecules on solid surfaces should allow one to create a new two-dimensional (2D) hybrid superconducting system, but its possibility has never been fully investigated so far. Here we report the fabrication of highly ordered 2D hybrid superconductors based on indium atomic layers on silicon surfaces and magnetic metal-phthalocyanines (MPc) and clarify their detailed structural, superconducting and magnetic properties. Our primary findings include a substantial controllability of the superconducting transition temperatures ($T_c$) through substitution of central metal ions (M = Cu, Fe, Mn) of the molecules. This is attributed to charge transfers between the magnetic molecules and the superconducting layers and to different degrees of exchange coupling between them, which originates from anisotropic distributions of the relevant *d*-orbitals. The present study opens a route for designing and creating exotic 2D superconductors with an atomic-scale precision.**

Superconductivity and magnetism have traditionally been regarded as conflicting phenomena. Since the former in the conventional form retains the time reversal symmetry, it is adversely affected in general by the latter that breaks this symmetry. However, studies during the past decades have revealed that superconductivity can coexist with magnetism in certain conditions and may even be enhanced by it[1-5]. It is also possible to create an exotic state of matter, topological superconductivity, based on the combination of conventional superconductivity and magnetic exchange interaction or field together with the Rashba effect[6-8]. In this respect, the recent establishment of atomic-layer superconductors[9-16] and self-assembly of magnetic molecules[17-19], both on solid surfaces, offer an opportunity for creating an ideal two-dimensional (2D) hybrid system[20] (see Fig. 1a). Here the metal atomic layer on a semiconductor surface, technically called a metal-induced surface reconstruction[21], plays the role of a host 2D material. The ultimately small layer thickness allows its *macroscopic* properties to be tuned by an external perturbation from a guest adsorbate, which in this case are organic molecules. The usage of organic molecules is advantageous in terms of flexible and rational designing, potentially leading to i) fine adjustment of the adsorbate-substrate interaction and ii) manifestation of additional functions through a self-assembly process[22,23]. The concept might seem analogous to a previous study on self-assembled magnetic molecules on polycrystalline gold films in solution[24], but our approach should enable an unprecedentedly precise control of such a complex 2D system. The realisation of this prospect, however, requires detailed investigations on structural, electronic and magnetic properties and their influences on superconductivity. Particularly, information on the exchange coupling between the local magnetic moments of the molecules and the conduction electrons in the atomic layer will be crucial.

In this Article, we demonstrate the fabrication of hybrid 2D superconductors based on the Si(111)-($\sqrt{7}\times\sqrt{3}$)-In surface [referred to as ($\sqrt{7}\times\sqrt{3}$)-In][9-12,25-27] and metal-phthalocyanines (MPc, M = Cu, Fe, Mn)[28-33] and clarify their detailed properties by scanning tunnelling microscopy (STM), electron transport measurement, X-ray magnetic circular dichroism (XMCD) and *ab initio* calculation. All MPc molecules are found to form a highly ordered monolayer on the ($\sqrt{7}\times\sqrt{3}$)-In surface, locking their lattice orientations against the crystallographic directions of the substrate. Despite their nearly identical assembly structures, the MPc molecules affect the superconductivity of the ($\sqrt{7}\times\sqrt{3}$)-In surface in strikingly different manners; CuPc increases $T_c$ while FePc and MnPc suppress $T_c$ moderately and strongly, respectively. In case of MnPc, a resistance minimum is observed after disappearance of superconductivity, indicating the manifestation of the Kondo effect. Spin magnetic moments at the metal ions are unambiguously detected for all adsorbed MPc molecules by XMCD measurement. *Ab initio* calculations clarify not only the presence of spins at metal ions but also their different degrees of coupling with the indium layer. The suppression of $T_c$ observed for FePc and MnPc is attributed to strong exchange couplings between the molecular spins and the conduction electrons, the strengths of which are estimated from the decrease in $T_c$. In contrast, the observed increase in $T_c$ for CuPc is ascribed to a significant charge transfer from the substrate to the molecules, which is corroborated by *Ab initio* calculations.

## Atomic-layer superconductor and magnetic molecules under study

The ($\sqrt{7}\times\sqrt{3}$)-In surface was chosen as an atomic-layer superconductor because the emergence of superconductivity with a transition temperature $T_c \cong 3$ K has firmly been established by STM and electron transport measurements [9-12,27]. It consists of double indium atomic layers on a silicon surface, resembling In(100) planes that conform to a Si(111) substrate with a $\sqrt{7}\times\sqrt{3}$ periodicity[34,35]. This new periodicity leads to the backfolding of a free-electron-like indium *sp* bands, forming a butterfly-like Fermi surface[26]. MPc molecules coordinated with a metal ion at centre are characterised by $D_{4h}$ symmetry. With a transition metal ion, they often exhibit magnetism because of partially occupied *d*-orbitals and finite spin magnetic moments. This allows their magnetism to be tuned through substitution of metal ions. In the present study, we used CuPc, FePc and MnPc, which possess spin states $S =1/2$, 1 and 3/2 in the gas phase, respectively[28,29,31,32] (for schematic spin configurations, see Fig. 1b).

## Molecular assembly structures on the indium atomic layer on silicon

First, the assembly structures of MPc molecules on the ($\sqrt{7}\times\sqrt{3}$)-In surface were investigated in detail with a low-temperature (LT) STM at 4.6 K. Figure 1d shows a representative STM image of the pristine (√7×√3)-In surface. Atomic structures with parallel rows running in the $[11\bar{2}]$ direction of the silicon substrate are clearly resolved, where the √7×√3 unit cells are indicated by the yellow parallelograms. The structural modulation along the $[\bar{1}10]$ direction has not been reported previously, but this was always found in our experiments regardless of the doping type of the silicon substrate. All MPc molecules were found to be adsorbed in the in-plane geometry, which was also confirmed by X-ray absorption spectroscopy (XAS) of N $K$-edge (see Supplementary Information). The molecular growth was nearly in the layer-by-layer mode at least within a few monolayers regime.

Figures 1e-1g shows STM images of the surface covered with monolayers of CuPc, FePc and MnPc. Closely packed molecular square lattices were found in all cases, which are reminiscent of the assemblies of MPc on metal surfaces[36,37]. The principal axes of the observed lattices were rotated by 45° from the parallel rows of the (√7×√3)-In surface ($[11\bar{2}]$ direction). Our detailed analysis revealed that the √2×√2 unit cell of the MPc monolayers matches the 3×2 unit cell of the underlying (√7×√3)-In surface within an error of ~3% (Fig. 1c; see Supplementary Information). This good commensurability leads to the orientation locking and a slight distortion of the square molecular lattice. However, the adsorption site of the molecule was not identified because of frequent phase shifts observed in the (√7×√3)-In surface. The areal density $\sigma$ of MPc molecules was determined to be 0.522 nm$^{-2}$. Since $\sigma$ = 18.8 nm$^{-2}$ for the double-layer indium atoms, the fractional concentration of the molecules against indium atoms is 2.8%, which will be used for estimation of the exchange coupling below.

**Superconducting transitions of the indium atomic layer-molecule hybrids**

Superconductivity of the samples was observed through electron transport measurements, which were performed under the ultrahigh vacuum (UHV) condition to avoid surface contamination and degradation[10,27] (see Methods). After the first transport measurement of the pristine (√7×√3)-In surface down to 1.6 K, one of the three MPc molecules was adsorbed to a sub-ML coverage and the measurement was taken again. This process was repeated several times until the molecule coverage $\theta$ reached about 2 ML.

Figures 2a and 2b depict the results for CuPc and FePc overlayers with different coverages $\theta$, where sheet resistance $R_{2D}$ was plotted as a function of temperature $T$. They exhibit sharp superconducting transitions at 2.9-3.2 K for $0 \leq \theta \leq 2.0$ ML, indicating that the (√7×√3)-In structure was well preserved under the molecular layers (regarding the determination of $T_c$, see Supplementary Information). Notably, the adsorption of CuPc led to increase in $T_c$ from the pristine value of 3.05 K to 3.20 K when $\theta$ was increased to 0.8 ML, corresponding to a 5% increase in $T_c$. This change was nearly offset by an additional increase in $\theta$ to 2.0 ML. In case of FePc, the adsorption induced a lowering of $T_c$ that amounts to 5% at $\theta$ = 1.8 ML. In contrast, this robustness of superconductivity was not observed for the MnPc overlayer. Figure 2c shows similar experimental runs using MnPc for $0 \leq$ $\theta \leq 1.8$ ML. Adsorption of 0.3 ML lowered $T_c$ by ~30% and the superconducting transition was not observed for $\theta \geq 0.6$ ML down to 1.6 K. These experiments were repeated for each molecular species, which gave the same trends as described above. The results are summarised in Fig. 2d where the changes in $T_c$ are plotted as a function of $\theta$. We note that, for MnPc, the temperature dependence of $R_{2D}$ showed a plateau-like region around 3 K for $\theta = 0.6$ ML and a local minimum around 4-5 K for $\theta \geq 1.2$ ML (see Fig. 2e). This indicates the occurrence of the Kondo effect due to the spin magnetic moments of MnPc[24]. Fitting a theoretical form to the data allowed us to estimate the Kondo temperature $T_K = 12 \pm 2$ K (see Supplementary Information).

The characteristic dependence of $T_c$ found above should be ascribed to distinctive electronic and magnetic states of individual MPcs. Generally, when organic molecules are adsorbed on a substrate, charge transfer occurs between them due to the difference in their chemical potentials. Since this plays the role of carrier (electron or hole) doping to the indium layer, the density of states at the Fermi level, $\rho(E_F)$, varies if $\rho$ is energy dependent. Then $T_c$ is modulated according to the Bardeen-Cooper-Schrieffer (BCS) theory[38]. Aside from this effect, the presence of local spins suppresses superconductivity in general due to their exchange coupling to the conduction electrons[38-40]. However, spins of MPc molecules are often reduced in size and can even be quenched by adsorption through substrate-induced charge transfer and electronic reconfiguration within the $d$-orbitals[31,32]. In the following, these problems are studied by XMCD and *ab initio* calculations.

**Spin magnetic moment of adsorbed molecules: XMCD study**

Spin magnetic moment due to the $d$-orbitals of transition metal can be directly measured by XMCD using a circularly polarised X-ray beam [17,18,41-45]. XMCD signal is defined as the difference between two XAS intensities measured with photon helicities parallel and antiparallel to the applied magnetic field. The experiments were performed at $T = 5$ K and under magnetic fields of $B = \pm 5$ T with a beam incident angle of 55°. The coverage $\theta$ of MPc molecules was set to 1 ML. Figures 3a-3c show XAS and XMCD spectra taken at the $L$-edges of Cu, Fe and Mn. For all cases, clear XMCD signals were detected, verifying the presence of spins at these metal ions. Analysis based on the XMCD sum rule allowed us to determine the effective spin magnetic moment $<m_s^{\text{eff}}>$[42,45]. Although $<m_s^{\text{eff}}>$ generally includes a contribution from spin dipole moment, this is negligibly small for the incident angle of 55° and $D_{4h}$ symmetry of the molecules. Hence we simply identify $<m_s^{\text{eff}}>$ with spin magnetic moment $<m_s>$. The $<m_s>$ values were determined to be $0.33 \pm 0.13$ $\mu_B$ for CuPc, $0.51 \pm 0.22$ $\mu_B$ for FePc and $0.97 \pm 0.14$ $\mu_B$ for MnPc, where $\mu_B$ is the Bohr magneton (Table 1; see Supplementary Information). These spins are likely to be in the paramagnetic state due to a weak intermolecular coupling below 1 ML coverage[36]. They are reduced in size by a factor of the Brillouin function $B_S(x)$, where $x \equiv gS\mu_B B/k_B T$ ($g = 2$, $k_B$: the Boltzmann constant). Assuming that $S = 1/2$ for CuPc/FePc and $S = 1$ for MnPc based on the *ab initio* calculations (see below), $B_{S=1/2}$ ($x = 0.67$) =

0.59 for the former and $B_{S=1}(x =1.34) = 0.70$ for the latter. These values are used for comparison with the *ab initio* results in the following.

**Electronic and magnetic states of adsorbed molecules: *ab initio* study**

To clarify the electronic and magnetic states of MPc molecules adsorbed on the ($\sqrt{7}\times\sqrt{3}$)-In surface, we performed *ab initio* calculations (see Methods). The result confirms that all MPcs retain spin magnetic moments $m_s$ after adsorption; when projected onto the *d*-orbitals of the metal ions, they are 0.49 μ$_B$ (CuPc), 1.19 μ$_B$ (FePc) and 2.37 μ$_B$ (MnPc) (see Table 1). These values correspond to $S \cong$ 1/2 for CuPc/FePc and $S \cong 1$ for MnPc as mentioned earlier. Compared to the *ab initio* results, the $<m_s>$ values determined by XMCD are smaller by factors of 0.67±0.27, 0.43±0.18 and 0.41±0.06, respectively. These reduction factors are consistent with the Brillouin function $B_S(x) = 0.59$ for CuPc/MnPc, while a discrepancy from $B_S(x) = 0.70$ is noticeable for MnPc. This deviation may be ascribed to the underscreened Kondo effect[46]; a local spin larger than $S = 1/2$ can be *partially* Kondo-screened by many-body interaction with conduction electrons through certain channels[40,47,48]. Then the spin is reduced in size below the relevant Kondo temperature $T_K$, but a residual local spin can persist even below $T_K$. We note that $T = 5$ K for the XMCD measurement is lower than $T_K = 12\pm2$ K deduced from the transport measurement (Fig. 2e), suggesting that the spin of MnPc is indeed partially screened. Another possible reason for the deviation is underestimation of $<m_s>$ due to a partial overlapping of the $L_{3,2}$ peaks for MnPc[49](see Supplementary Information).

Our *ab initio* calculations also elucidate the origin of the observed distinctive influences of the MPc molecules on superconductivity. Figures 3d-3f show the spatial spin distributions of CuPc, FePc and MnPc adsorbed on the ($\sqrt{7}\times\sqrt{3}$)-In surface. For CuPc, the spin distribution is confined within the molecule because the orbital responsible for the spin, $d_{x^2-y^2}$, extends in the in-plane direction. Correspondingly, spin magnetic moment projected on the In *p*-orbitals, $m_{s\_In}$, is found to be 0.000 μ$_B$. In contrast, it penetrates into the indium layer for FePc and MnPc because the relevant orbitals, $d_{xz}$, $d_{yz}$ and $d_{z^2}$, extend also in the out-of-plane direction. This leads to $m_{s\_In}= -0.034$ μ$_B$ for FePc and $m_{s\_In}= -0.035$ μ$_B$ for MnPc, both of which are localised under the molecules. Here the negative sign indicates the antiferromagnetic coupling. Furthermore, the density of states projected on the *d*-orbitals show that $d_{x^2-y^2}$ of CuPc exhibits narrow peaks while significant energy broadenings are noticed for $d_{xz}$, $d_{yz}$, $d_{z^2}$ of FePc and MnPc (Figs. 3g-3i). These results show that the coupling of the spin-related orbitals of the molecule to the superconducting layer is weak for CuPc while strong for FePc and MnPc. This explains why $T_c$ was lowered only for the latter. The origin of the different behaviours between FePc and MnPc is not clear, but it may be attributed to the occurrence of the Kondo effect for MnPc; competition between superconductivity and the Kondo effect can lead to a strong suppression of the former[50].

Considering that the weak interaction of CuPc should leave the electron-phonon coupling in the indium layer nearly unaffected, the observed increase in $T_c$ for CuPc is ascribed to a carrier doping into the indium layer. Our calculations revealed that 1.61 electrons are transferred from the indium layer to a CuPc molecule (Table 1). Since the fractional concentration of CuPc molecules is 2.8% (see above), this corresponds to a 0.045 hole doping per indium atom. Together with the relatively large variation of $\rho(E)$ of the (√7×√3)-In surface around $E_F$[34], this can explain the enhancement of $T_c$. The *ab initio* calculations also showed charge transfers of 2.09 and 2.01 electrons per molecule in the same direction for FePc and MnPc, but their contributions must be dominated by the effect of exchange coupling described above.

## Estimation of the exchange coupling

The adsorbed MPc molecules are likely to be paramagnetic as mentioned above. Based on this assumption, we now estimate the strength of their exchange coupling with the underlying indium layers. According to the Abrikosov-Gorkov theory, the effect of magnetic impurities (local spins) on superconductivity can be measured by the pair-breaking parameter $2\alpha$[38]:

$$2\alpha \cong \frac{xJ^2}{E_F} , \qquad (1)$$

where $x$ is the fractional impurity concentration, $J$ is the exchange coupling between the magnetic impurity and the host electrons ($J > 0$ for ferromagnetic and $J < 0$ for antiferromagnetic coupling). Then the $T_c$ lowering is given by the following equation:

$$k_B(T_{c0} - T_c) = \frac{\pi\alpha}{4} , \qquad (2)$$

where $T_{c0}$ is the $T_c$ value in the absence of magnetic impurities. The transport data in Figs. 2b and 2c show that $T_{c0} - T_c = 0.08$ K for 0.9 ML of FePc ($x = 0.0252$) and $T_{c0} - T_c = 1.0$ K for 0.3 ML of MnPc ($x = 0.0084$). Together with $E_F = 6.9$ eV for the (√7×√3)-In surface[26], substitution of these parameters into Eqs. (1) and (2) gives $|J| = 0.070$ eV for FePc and $|J| = 0.46$ eV for MnPc. The increase in $T_c$ precludes this analysis for CuPc, but $|J|$ should be smaller than these values according to the transport data and the *ab initio* results. We note that these estimations are susceptible to errors due to charge transfer effects and the possible Kondo effect for MnPc. The data for MnPc may also have been influenced by phase separation due to molecular island formation at $\theta = 0.3$ ML.

Our experiments have revealed that the magnetic moments of FePc, which are likely to be in the paramagnetic state, can coexist with superconductivity. In this situation, the spins are randomly oriented and the effective Zeeman field $B_{\text{eff}}$ experienced by Cooper pairs becomes zero. This is why the effect of magnetic impurities on superconductivity is relatively limited. However, if the spins are aligned by external magnetic field $B_{\text{ext}}$, $|B_{\text{eff}}|$ becomes sufficiently large to destroy superconductivity (the Pauli pair-breaking effect). A simple estimation using a relation $B_{\text{eff}} = xJ<S>/g\mu_B$ [51] and parameters $x = 0.028$, $|J| = 0.070$ eV, $<S> = 1/2$ for FePc gives $|B_{\text{eff}}| = 8.4$ T. This value is larger than $B_{c2}^{\text{Pauli}} \equiv \Delta_0/\sqrt{2}\mu_B \cong 5$ T ($\Delta_0 = 1.76 k_B T_c$: BCS energy gap) where $B_{c2}^{\text{Pauli}}$ is the critical magnetic field due to the Pauli pair-breaking effect. In case of in-plane magnetic field, superconductivity is broken not by the usual orbital pair-breaking effect but by this mechanism. Furthermore, superconductivity suppressed this way can be revived if the external field $B_{\text{ext}}$ are adjusted to compensate $B_{\text{eff}}$, i.e., if $B_{\text{eff}} + B_{\text{ext}} = 0$ is satisfied (the Jaccarino-Peter effect)[3,51]. This is possible if the relevant exchange interaction is antiferromagnetic ($J < 0$), which is supported by our *ab initio* calculations as mentioned above. Therefore, superconductivity of this 2D hybrid system will be controlled in a non-trivial manner by external magnetic field.

**Conclusion**

In this Article, we have demonstrated the fabrication of 2D hybrid superconductors based on the indium atomic layer on silicon and magnetic MPc molecules, the properties of which have been clarified through multiple experimental and theoretical techniques. The unambiguous confirmation of the spin magnetic moments of MPc and their strikingly different influences on $T_c$ have shown the feasibility of controlling *macroscopic* superconducting properties in a reproducible manner. Finally, we remark that the Rashba effect due to the spatial inversion symmetry breaking at a surface will play an important role if heavier elements are used instead of indium. This effect can substantially enhance the robustness of 2D superconductivity against magnetism since electrons comprising a Cooper pair become spin-polarised even in the absence of magnetic interaction[52]. Indeed, a large spin splitting of the Fermi surface and extremely high in-plane critical magnetic field originating from the Rashba effect have already been observed for this type of atomic-layer superconductors[14,15]. This strengthens the compatibility of superconductivity and magnetism and makes the creation of topological superconductivity more realistic[6,20]. Thus the present 2D hybrid system will provide a basic platform to explore such exotic superconductors.

**Methods**

Non-doped commercial silicon wafers (resistivity $\rho > 1000$ $\Omega$cm) were used as sample substrates for transport measurements to avoid the contribution of substrate conduction, while N-type doped wafers were used for LT-STM and XMCD measurements ($\rho < 0.01$ $\Omega$cm for LT-STM and $\rho = 1-3$ $\Omega$ cm for XMCD). The substrate was first heated above 1250 °C to obtain a clean Si(111) surface. Following thermal deposition of a small amount of indium, it was annealed around 300 °C to prepare the ($\sqrt{7}\times\sqrt{3}$)-In surface[10,27]. The surface phase was confirmed by LEED and/or STM at room temperature, and the preparation condition was optimised so that large domains with a small density of defects were obtained. The MPc molecules with purities better than 99.99% (CuPc), 96% (FePc) and 99% (MnPc) were further purified by being degassed for several hours in UHV. The molecule coverage was estimated by STM observation of the sample surface for transport measurements and by combination of quartz balance monitoring and XAS signals of N K-edge for XMCD measurements (see Supplementary Information).

All measurements were performed under the UHV condition (~1×10$^{-10}$ mbar). The molecular assembly structures were studied in detail using a LT-STM equipped with a liquid He cryostat. Transport experiments were performed in a home-built UHV apparatus[10,27]. A shadow mask was used to define the surface transport region with seamlessly connected leads and electrodes; the surface indium layer of the unmasked region was etched by argon ion beam. The sample was then transferred to the transport measurement unit where four gold-coated spring probes were tightly pressed onto the electrode regions.

XMCD measurements were performed at the beam line BL4B of UVSOR-III in Institute for Molecular Science, Japan. The experiments were conducted in a UHV system equipped with a superconducting split-coil magnet and a liquid He cryostat[44]. The XAS spectra recorded in a total-electron-yield mode were scaled by an incident X-ray intensity measured with a gold mesh placed upstream of the sample. They were further normalised by the intensity at an energy ~20 eV below the L-edge peaks, followed by tilt correction and averaging over 20-80 runs. Background signals from the substrate, identified with the XAS spectra for a clean (√7×√3)-In surface, were subtracted from the data. Although the samples were exposed to X-rays for several hours during the repeated measurements, no sample damaging was observed. The effective spin magnetic moments $<m_s^{\text{eff}}>$ (identified with $<m_s>$ here) were calculated using the XMCD sum rule (see Supplementary Information) and were further corrected by dividing them by the polarisation degree of the beam (= 60%). The errors in $<m_s^{\text{eff}}>$ were estimated from the standard deviations of the spectra in the averaging process.

*Ab initio* calculations were performed based on the density functional theory using the plane wave–based Vienna *ab initio* simulation package (VASP) with the projected augmented wave (PAW) method[48]. To treat the *d*-electron states in the Fe atom, we have employed the LDA+U method with a setting of $U - J = 1.0$ eV[53]. For the modelling of the (√7×√3)-In surface, a slab of the double indium layer[34,35] and eight Si(111) layers terminated with hydrogen at the backside was employed. The adsorption site for a molecule was chosen to be the on-top site, which was found to be of the lowest energy for MnPc. The in-plane mirror axis of the MPc molecule was set at an angle of 15° against the $[11\bar{2}]$ direction based on the STM observations (see Supplementary Information). The positions of all atoms were optimised until the forces on individual atoms were less than 0.02 eV/Å. Because of the large dimensions of the supercell, the Brillouin zone was sampled only at Γ point. The charge transfer between the MPc molecule and the substrate was deduced from the Bader analysis of the *Ab initio* result.

**Acknowledgments**

The following financial supports are acknowledged: JSPS Kakenhi Grant No. 25247053 (T.U.), World Premier International Research Center (WPI) Initiative on Materials Nanoarchitechtonics (S.Y., S.V., T.N. and T.U.), JSPS KAKENI Grant No. 25871114 and 15K17465 and MEXT KAKENHI Grant No. 25110008 and 26102017 (E.M.). The XMCD work was partly supported by Nanotechnology Platform Program (Molecule and Material Synthesis) of MEXT, Japan (Proposal Nos. S-13-MS-1058, S-14-MS-2011, S-15-MS-2034). The calculations were performed at the computer facilities of the Institute of Solid State Physics (ISSP Super Computer Center, University of Tokyo) and RIKEN (HOKUSAI GreatWave). S.Y., T.U.,Y.T, and T.Y. thank Ms. M. Uozumi for her technical helps during the XMCD measurements. E.M. and T.U. thank R. Arafune for fruitful discussions and for his technical support.

**Author contributions**

S.Y and T.U. conceived and designed the research project. Experiments and analysis of data were performed by: S.Y., S.V., P.M. and T.U. (STM), S.Y. (electron transport), S.Y. and Y.T. (XMCD). E.M. performed theoretical calculations. T.Y., S.W. and T.N. supervised the work. S.Y., E.M. and T.U. wrote the paper. All authors discussed the results and commented on the manuscript.

**Additional Information**

Supplementary information is available in the online version of the paper. Reprints and permission information is available online at http://www.nature.com/reprints. Correspondence and requests for materials should be addressed to T.U.

**Competing financial interests**

The authors declare no competing financial interests.

**Figure captions**

**Figure 1 | Concept and assembly structures of 2D hybrid superconductors. a,** Concept of a 2D hybrid superconductor consisting of the metal atomic layer (blue spheres) on a semiconductor (grey spheres) and self-assembled magnetic organic molecules (on top). Charge transfer and local spins due to the presence of organic molecules modify the *macroscopic* superconducting properties of the atomic layer. **b,** Schematic diagrams of the *d*-orbitals of the coordinated metal ions of MPc (M = Cu, Fe, Mn) in the gas phase. CuPc, FePc and MnPc have spin states of $S = 1/2$, $S = 1$ and $S = 3/2$, respectively. The up/down arrows indicate the spin directions of electrons at individual orbitals. **c,** Schematic view of the square lattice of the MPc assembly and the commensurate relation against the ($\sqrt{7}\times\sqrt{3}$)-In surface. The $\sqrt{2}\times\sqrt{2}$ unit cell of the MPc monolayers (blue dashed square) is equivalent to the 3×2 unit cell of the ($\sqrt{7}\times\sqrt{3}$)-In surface (yellow parallelograms). **d-g,** STM images of the pristine ($\sqrt{7}\times\sqrt{3}$)-In surface (**d**) and the monolayers of CuPc (**e**), FePc (**f**) and MnPc (**g**) on the ($\sqrt{7}\times\sqrt{3}$)-In surface. Image size: 12 nm ×12 nm, sample bias voltage: $V_s$ = 0.25 V (**d**), -2 V (**e**), -2 V (**f**) and 2.3 V (**g**), temperature: $T$ = 4.6 K. The arrows indicate $[11\bar{2}]$ and $[\bar{1}10]$ directions of the Si(111) substrate. Atomic structures with parallel rows running in the $[11\bar{2}]$ direction are visible in **d**, while the principal axes of the molecular lattices are rotated by 45° against these directions in **e-g**. The yellow parallelograms and blue dashed squares correspond to those depicted in **c**.

**Figure 2 | Electron transport measurements of the ($\sqrt{7}\times\sqrt{3}$)-In surfaces with MPc overlayers. a-c,** Temperature dependences of the 2D resistivity of the ($\sqrt{7}\times\sqrt{3}$)-In surface with overlayers of CuPc (**a**), FePc (**b**) and MnPc (**c**) for different molecular coverages ($0 \leq \theta \leq 2$ ML). CuPc enhances $T_c$ while FePc and MnPc suppresses $T_c$ moderately and strongly, respectively. The results are summarised in **d**. **e,** Magnified display of the MnPc data in **c** for $\theta \geq 0.6$ ML where superconductivity is quenched. A plateau (0.6 ML) and local minima (1.2, 1.8 ML) of resistance are noticeable, indicating the occurrence of the Kondo effect. The red line is a fit to the data for $\theta \geq 1.8$ ML by an empirical form[54,55] (see Supplementary Information).

**Figure 3 | Electronic and magnetic states of MPc molecules on the ($\sqrt{7}\times\sqrt{3}$)-In surfaces. a-c,** XAS/XMCD spectra of monolayers of CuPc (**a**), FePc (**b**) and MnPc (**c**) on the ($\sqrt{7}\times\sqrt{3}$)-In surface. The red and blue curves are XAS spectra taken with circularly polarised light having a fixed photon helicity under magnetic fields of ±5 T. The green curves are XMCD spectra defined as the difference between the two XAS spectra, the intensities of which are multiplied by a factor of 2 for display. Clear XMCD signals are detected for all cases, revealing the presence of spin magnetic moments. The experiments were performed at 5 K. **d-f,** *Ab initio* calculations of the spatial spin distributions of CuPc (**d**), FePc (**e**) and MnPc (**f**) adsorbed on the ($\sqrt{7}\times\sqrt{3}$)-In surface. The spin of CuPc is confined within the molecule while it penetrates into the indium layer for FePc and MnPc. The figures were rendered

by VESTA [K. Momma and F. Izumi, *J. of Appl. Crystal.* **44**, 1272 (2011)]. The isosurface value was set to 0.001 $e/a_B^3$ ($a_B$: Bohr radius). The yellow and light-blue colours correspond to majority and minority spins, respectively. **g-i,** *Ab initio* calculations of the spin-polarised density of states projected on the *d*-orbitals of CuPc (**d**), FePc (**e**) and MnPc (**f**) adsorbed on the (√7×√3)-In surface. The energy is measured relative to $E_F$. The data corresponding to the spin-related orbitals are displayed with vertically filled lines (CuPc: $d_{x^2-y^2}$, FePc: $d_{xz}, d_{yz}, d_{z^2}$, MnPc: $d_{xz}, d_{yz}, d_{z^2}, d_{xy}$). The $d_{x^2-y^2}$ orbital exhibits a narrow peak while significant energy broadenings are noticed for $d_{xz}, d_{yz}, d_{z^2}$ (especially $d_{z^2}$), showing the different degrees of coupling to the indium layers.

**Table 1 | Spin magnetic moments and charge transfers for MPc monolayers on the (√7×√3)-In surfaces.** The spin magnetic moments determined by XMCD, $<m_s>$, spin magnetic moments projected onto the *d*-orbitals of the metal ions obtained by *ab initio* calculations, $m_s$, and the ratio $<m_s>/m_s$ are displayed for CuPc, FePc and MnPc. The Brillouin functions $B_S(x)$, where $x \equiv gS\mu_B B/k_B T$, were calculated assuming that $S = 1/2$ for CuPc/FePc and $S = 1$ for MnPc from the *ab initio* result. $B = 5$ T and $T = 5$ K were chosen from the XMCD experiment. Electron transfers ΔQ from the indium layer to MPc per molecule, which were deduced from the Bader analysis of the *ab initio* result, are also shown in the unit of elemental charge *e*.

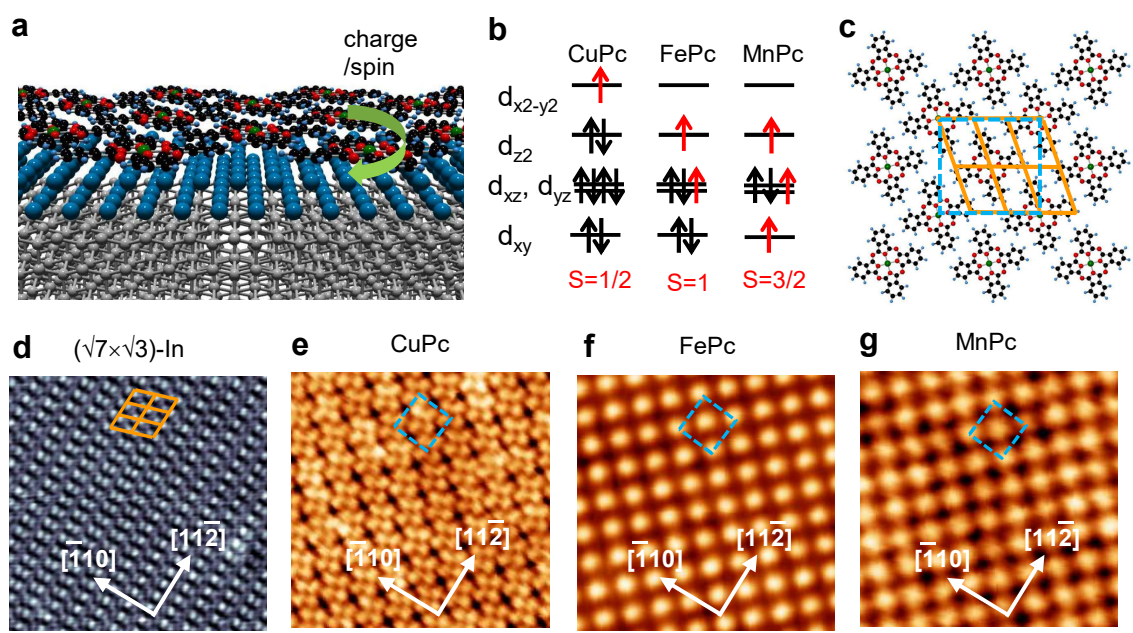

Figure 1

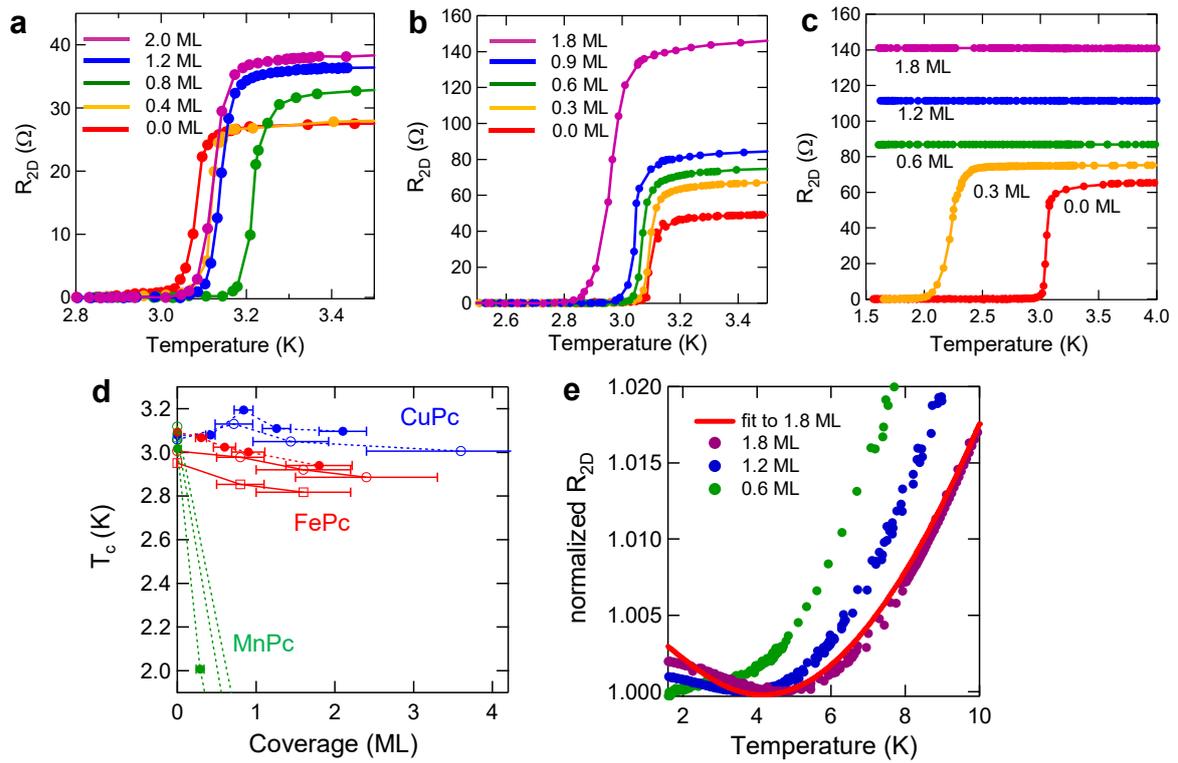

Figure 2

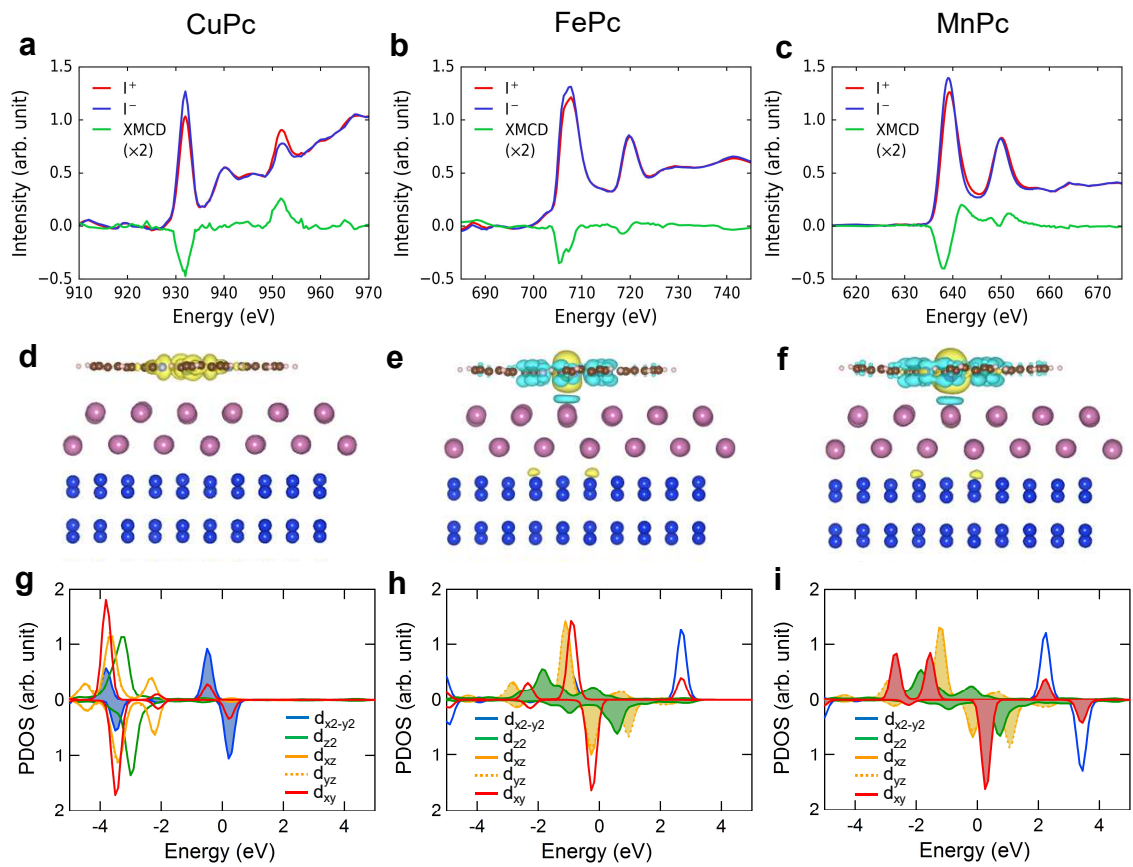

Figure 3

|  | CuPc | FePc | MnPc |
|---|---|---|---|
| XMCD, $<m_s>$ ($\mu_B$) | 0.33 ± 0.13 | 0.51 ± 0.22 | 0.97 ± 0.14 |
| *Ab initio*, $m_s$ ($\mu_B$) | 0.49 | 1.19 | 2.37 |
| $\frac{(XMCD, <m_s>)}{(Ab\ initio,\ m_s)}$ | 0.67 ± 0.27 | 0.43 ± 0.18 | 0.41 ± 0.06 |
| $B_s(x)$ | 0.59 | 0.59 | 0.70 |
| *Ab initio*, $\Delta Q$ ($e$) | 1.61 | 2.09 | 2.01 |

Table 1

# Supplementary Information for "Two-dimensional atomic layer-molecule hybrid superconductors with controllable exchange coupling"

Shunsuke Yoshizawa[1], Emi Minamitani[2], Saranyan Vijayaraghavan[1†], Puneet Mishra[1], Yasumasa Takagi[3], Toshihiko Yokoyama[3], Satoshi Watanabe[2], Tomonobu Nakayama[1] and Takashi Uchihashi[1*]

[1]International Centre for Materials Nanoarchitectonics (MANA), National Institute for Materials Science, 1-1, Namiki, Tsukuba, Ibaraki 305-0044, Japan

[2] Department of Materials Engineering, the University of Tokyo, 7-3-1, Hongo, Bunkyo-ku, Tokyo 113-8656, Japan

[3]Institute for Molecular Science, Myoudaiji Campus, 38 Nishigo-Naka, Myodaiji, Okazaki, Aichi 444-8585, Japan

†Present Address: CSIR-Central Electrochemical Research Institute, India
*email: UCHIHASHI.Takashi@nims.go.jp

This file contains the following sections.

1. Determination of the assembly structures of MPc monolayers on the Si(111)-(√7×√3)-In surface

2. Determination of the superconducting transition temperature $T_c$ and the Kondo temperature $T_K$

3. XAS/XMCD measurements on MPc molecules on the Si(111)-(√7×√3)-In surface

# 1. Determination of the assembly structures of MPc monolayers on the Si(111)-(√7×√3)-In surface.

To clarify the assembly structures of MPc monolayers on the (√7×√3)-In surface, STM observations were performed at 4.6 K using a LT STM. In the following, the experimental result and the analysis are described in detail for CuPc. Although the sample bias voltages ($V_s$) needed to obtain clear STM images were different for the (√7×√3)-In surface and CuPc, comparison of Figs. S1a and S1b taken on the same area allowed us to determine the lattice orientation and the rotational direction of the individual molecules of CuPc monolayers. It was found that the principal axes of the molecular lattice were rotated by 45° from the $[11\bar{2}]$ direction of the Si(111) surface, along which the parallel rows of the (√7×√3)-In surface run (see also Fig. 1d). Since the molecules are closely packed, the centre-lobe direction (in-plane mirror axis of CuPc) is rotated by an angle of ~30° from one of the principal axes of the lattice. Consequently, it makes an angle of ~15° against the $[11\bar{2}]$ direction. This observation was always the case as far as the molecular lattice was confined within a single domain region of the (√7×√3)-In surface. This suggests that the orientation of the molecular lattice is locked due to a good commensurability with the underlayer. Indeed, as schematically shown in Fig. S1c, the 3×2 unit cell of the (√7×√3)-In is equivalent to a rectangle with side lengths of 1.995 nm in the $[11\bar{2}]$ direction ($\equiv a_{\text{In}[11\bar{2}]}$) and 1.920 nm in the $[\bar{1}10]$ direction ($\equiv a_{\text{In}[\bar{1}10]}$). This rectangle is very close in size to the √2×√2 unit cell of the observed molecular lattice (approximately 1.960 nm ×1.960 nm).

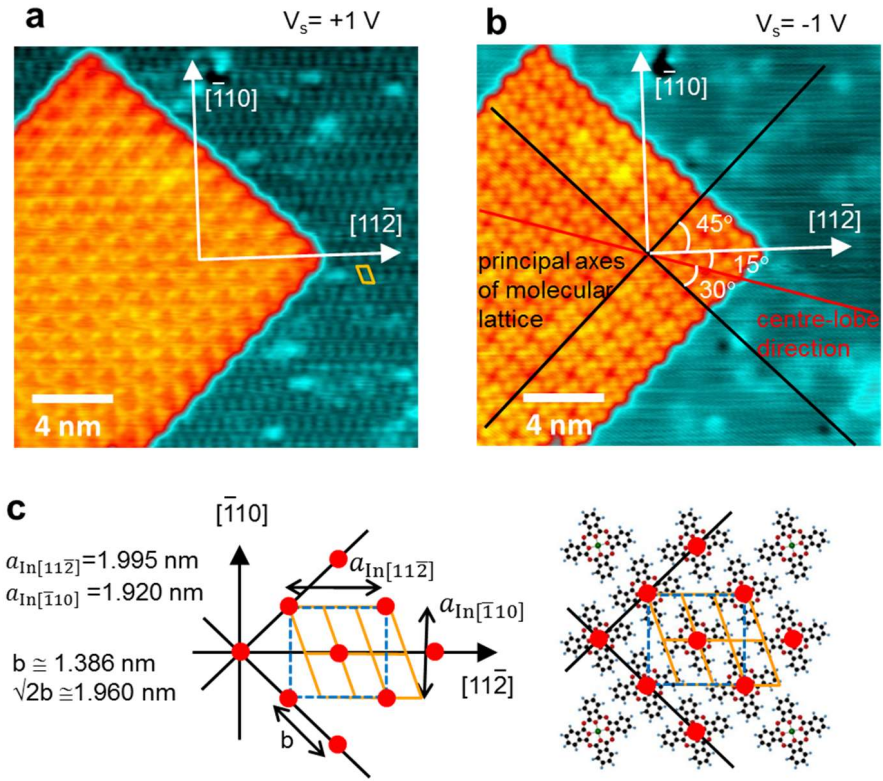

**Figure S1** | STM images and schematic diagram of a CuPc molecular lattice assembled on the (√7×√3)-In surface. **a, b** STM images of a monolayer CuPc lattice on the (√7×√3)-In surface taken at 4.6 K (**a**: $V_s$ = +1 V, **b**: $V_s$ = -1 V). **c** Relation between the CuPC lattice and the (√7×√3)-In surface. Red circles: locations of the CuPc molecules. Yellow lines: the 3×2 unit cell of the (√7×√3)-In. Blue dotted rectangule: the √2×√2 unit cell of CuPc lattices.

More detailed analysis on STM images allowed us to confirm that the commensurability was very good. Figure S2a shows an STM image where both CuPc monolayer islands and the (√7×√3)-In substrate are exposed on the surface. The fast Fourier transform (FFT) patterns of the former and the latter regions are displayed in Figs. S2b and S2c, respectively. The dashed lines indicate the unit cells in the reciprocal space while the blue arrows the unit vectors. Generally, STM images and their FFT patters are deformed due to an error in expansion coefficient calibration of the piezoelectric scanner and to a shear drift caused by thermal drift and creep. However, one can determine the precise periodicity of the CuPc lattice by utilizing the simultaneously observed (√7×√3)-In surface as a reference.

Figure S2d shows a unit cell in the real space with unit vectors $\boldsymbol{a}_1 = (a_{1x}, a_{1y})$ and $\boldsymbol{a}_2 = (a_{2x}, a_{2y})$, while Fig. S2e the corresponding unit cell in the reciprocal space with unit vectors $\boldsymbol{b}_1 = (b_{1x}, b_{1y})$ and $\boldsymbol{b}_2 = (b_{2x}, b_{2y})$. Suppose the unit vectors $\boldsymbol{a}_1$ and $\boldsymbol{a}_2$ are transformed into $\boldsymbol{a'}_1$ and $\boldsymbol{a'}_2$ through expansion by factors of *s, t* in the x and y directions (|*s*-1|<<1, |*t*-1|<<1), respectively, shear drift in the x direction by $\alpha a_{iy}$ (|$\alpha$|<<1, *i*=1,2) and rotation by an angle θ (|θ|<<1) (note that vertical shift in the y direction can be included in expansion in the y direction). Correspondingly, $b_1$ and $b_2$ are transformed into $b'_1$ and $b'_2$. To the first order of |*s*-1|, |*t*-1|, α and θ, one obtains

$$b'_{ix}/b_{ix} = 1/s - \theta(b_{iy}/b_{ix})$$
$$b'_{iy}/b_{iy} = 1/t - (\alpha - \theta)(b_{ix}/b_{iy}) \quad (1)$$
$$(i = 1,2).$$

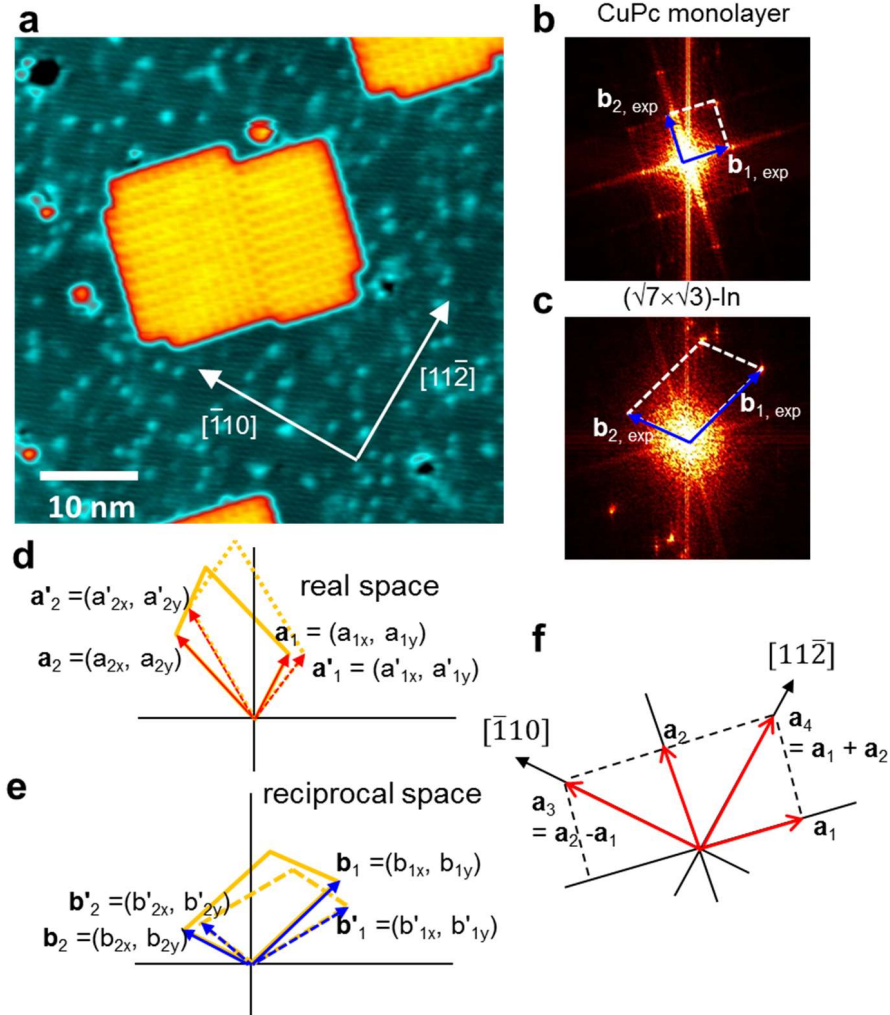

**Figure S2** | Determination of commensurate relation between the CuPc lattice and the (√7×√3)-In surface. **a,** STM image taken at 4.6 K ($V_s$ = +1 V). **b, c,** FFT patterns calculated from the CuPc monolayer region (**b**) and the (√7×√3)-In region (**c**) taken from **a**. **d, e** Schematic drawing of a unit cell and unit vectors before and after transformation in real space (**d**) and receprocal space (**e**). **f,** Relation between vectors $\boldsymbol{a}_1$, $\boldsymbol{a}_2$, $\boldsymbol{a}_3$, $\boldsymbol{a}_4$.

Hence, *s, t, α, θ* can be uniquely determined by identifying the experimentally obtained reciprocal unit vectors $b_{i,\,exp}$ with $b'_i$ (*i* =1,2).

Analysis using the FFT pattern in Fig. S2c led to *s* = 0.9888, *t* = 0.9964, α = -0.0190 and θ = 2.5567°. These parameters were used to correct the unit vectors $b_1$, $b_2$ of the CuPc lattice in Fig. S2b, which were in turn used to determine the unit vectors $a_1$, $a_2$ in the real space. If one defines $a_3 \equiv a_2 - a_1$, $a_4 \equiv a_2 + a_1$ (Fig. S2f), one finds |$a_3$|= 2.009 nm, |$a_4$|=1.977 nm. These values are in agreement with $a_{In[11\bar{2}]}$ = 1.995 nm and $a_{In[\bar{1}10]}$ =1.920 nm, respectively, within an error of 3%. Furthermore, the angle made by $a_3$ and $a_4$ is 89.0°, very close to the right angle. Therefore, the CuPc lattice is commensurate with the (√7×√3)-In surface. This means that the CuPc layer is slightly deformed from a perfect square lattice.

The same analysis was applied to other data on the monolayers of CuPc, FePc and MnPc. Within an error of ~3%, the obtained values of |$a_3$| and |$a_4$| were equal to $a_{In[11\bar{2}]}$ =1.995 nm and $a_{In[\bar{1}10]}$= 1.920 nm, respectively. This shows a good commensurability between the MPc molecule lattices and the (√7×√3)-In surface in general and thus rationalises the observed locking of the molecular lattice direction. The internal molecular structure was not imaged with STM for FePc and MnPc in our experiments (see Figs. 1f and 1g) because the in-plane $d_{x^2-y^2}$ orbital is located far above the Fermi level (see Figs. 3h and 3i) and is not involved in the STM imaging[1]. This precludes a direct determination of the orientation angles of individual FePc and MnPc molecules. Nevertheless, they should be identical to that of CuPc (~15° against the $[11\bar{2}]$ direction) considering the close packing and the same lattice constant of the molecule overlayers.

## 2. Determination of the superconducting transition temperature $T_c$ and the Kondo temperature $T_K$

Superconductivity in a 2D system generally exhibits a residual resistance near the Bardeen-Cooper-Schrieffer (BCS) condensation temperature, $T_c$, owing to thermal excitations of free vortices[2]. This makes it difficult to precisely determine $T_c$. Therefore, the following equation for the 2D resistivity $R_{2D}$ was used to deduce the accurate value of $T_c$ from the fitting of the experimental data[3,4]:

$$R_{2D}(T) = \left(G_{2D,n}(T) + G_{2D,s}(T)\right)^{-1}, \quad (2)$$

where $G_{2D,n}$ is the normal conductance and $G_{2D,s}$ is the contribution to the conductance due to the superconducting fluctuation effect above $T_c$. $G_{2D,n}$ was assumed to have the following form:

$$G_{2D,n}(T) = (R_n + aT^b)^{-1}, \quad (3)$$

where $R_n$ is the residual normal resistance at $T = 0$ and the temperature-dependent term, $aT^b$, expresses a power-law behaviour. $G_{2D,s}$ has a temperature dependence given by

$$G_{2D,s}(T) = \frac{1}{R_0}\frac{T}{T - T_c}, \qquad (4)$$

where the contributions from the Aslamazov-Larkin and the Maki-Thompson terms are included[5,6]. Precisely speaking, the latter term does not have this form because of the presence of a material-dependent parameter, but it was incorporated into $R_0$ to reduce the number of fitting parameters[4]. Figure S3a shows an example of the fitting results, from which $T_c$ = 3.02 K and $b$ = 1.83 were obtained. The value of the exponent $b$ is determined by the mechanism of conduction electron scattering, *e.g.* electron-phonon and electron-electron scatterings, and hence is not expected to change significantly by a molecular overlayer growth. This was indeed confirmed by our experiments where $b$ ~1.8 was always obtained from the analyses of the CuPc and FePc samples. Therefore, a constant value of $b$ = 1.8 was chosen for the following fitting procedure.

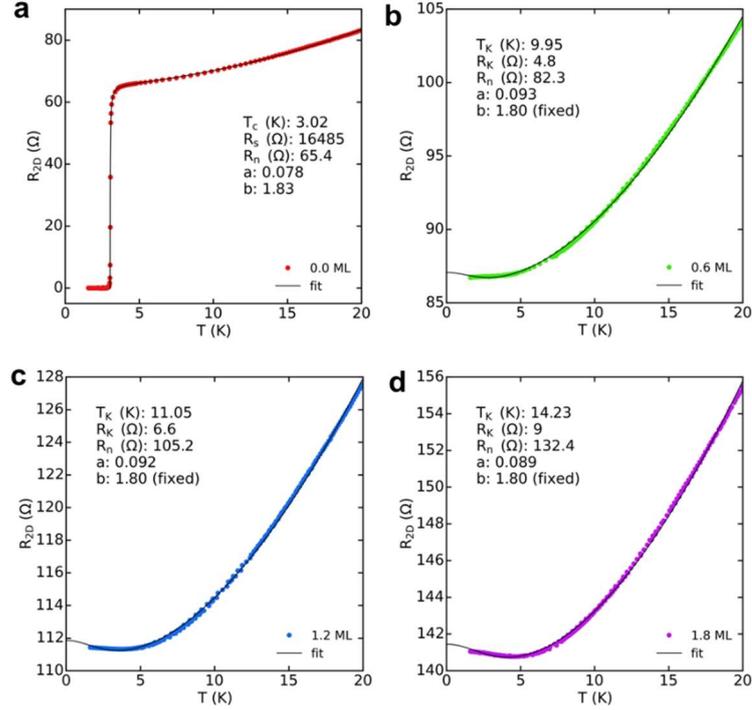

**Figure S3** | Determination of the superconducting transition temperature $T_c$ (**a**) and the Kondo temperature $T_K$ (**b-d**) from the temperature dependence of the resistance (**b**: $\theta$ = 0.6 ML, **c**: $\theta$ = 1.2 ML, **d**: $\theta$ =1.8 ML).

In case of the MnPc overlayer with $\theta \geq 0.6$ ML, superconductivity was found to be quenched and the signature of the Kondo effect was observed as explained in the main text. To determine the Kondo temperature $T_K$ by fitting, we used the following equation:

$$R_{2D}(T) = R_n + aT^b + R_K(T), \qquad (5)$$

where the term $aT^b$ is the same as in Eq. (3) and the exponent $b$ was chosen to be constant ($b$ =1.8) as mentioned above. Here $R_K(T)$ is the contribution to the resistance due to the Kondo scattering and has a following empirical form[7,8]:

$$R_K(T) = R_K(T=0)\left(\frac{T_K'^2}{T^2 + T_K'^2}\right)^s, \qquad (6)$$

where $T'_K = T_K/(2^{1/s} - 1)^{1/2}$ and $s = 0.225$ has been theoretically determined[9,10]. Figures S3b-S3d show the fitting result for MnPc with $\theta$ = 0.6, 1.2, 1.8 ML (the same data as in Fig. 2e), from which $T_K$ = 10.0, 11.0, 14.2 K was deduced, respectively. Thus $T_K$ can be estimated to be 12±2 K. Reasonably, $R_K(T=0)$ determined from the fitting was found to increase from 4.8 to 9.1 Ω as $\theta$ increased from 0.6 to 1.8 ML. We note that the experimental data tend to deviate from the theoretical curve at low temperatures as seen in Figs. S3b-S3d. This is not surprising considering that superconductivity may still occur in a nascent form below $T_c$ ~3 K even after it is suppressed by the local spin magnetic moments of MnPc.

## 3. XAS/XMCD measurements on MPc molecules on the Si(111)-(√7×√3)-In surface

Note: In this section, $\theta$ represents the incident angle of X-ray beam, not the coverage as used in the main text of the paper.

**XAS at N $K$-edge**

Figure S4a shows the N $K$-edge XAS spectra obtained for a CuPc monolayer on the (√7×√3)-In surface with $\theta$ = 0° and $\theta$ = 55°. They exhibit eight peaks similar to those reported for vanadium phthalocyanine[11]. The four peaks on the low- and high-energy sides are attributed to transitions from N 1$s$ to π* and σ* molecular orbitals, respectively, both of which have $p$ orbital characters [12,13]. At $\theta$ = 0° the peaks of π* orbitals are negligibly small, while those of σ* orbitals are enhanced. This means that the CuPc molecules were adsorbed in the in-plane geometry, because a dipolar transition from 1*s* to *p*$_z$-like π* orbitals is suppressed due to the selection rule when the X-ray beam is in the normal direction (see Fig. S4d). Figure S4b and S4c show the N *K*-edge XAS spectra for FePc and MnPc monolayers, displaying essentially the same spectral features as those of CuPc. This is consistent with the identical adsorption geometries of these MPcs revealed by STM. As shown in Figs. S4e and S4f, the intensity of each peak is approximately proportional to the molecular coverage. The nearly equal XAS intensities for the three MPc monolayers (Figs. S4a-S4c) demonstrate a precise coverage control in the XAS/XMCD experiments.

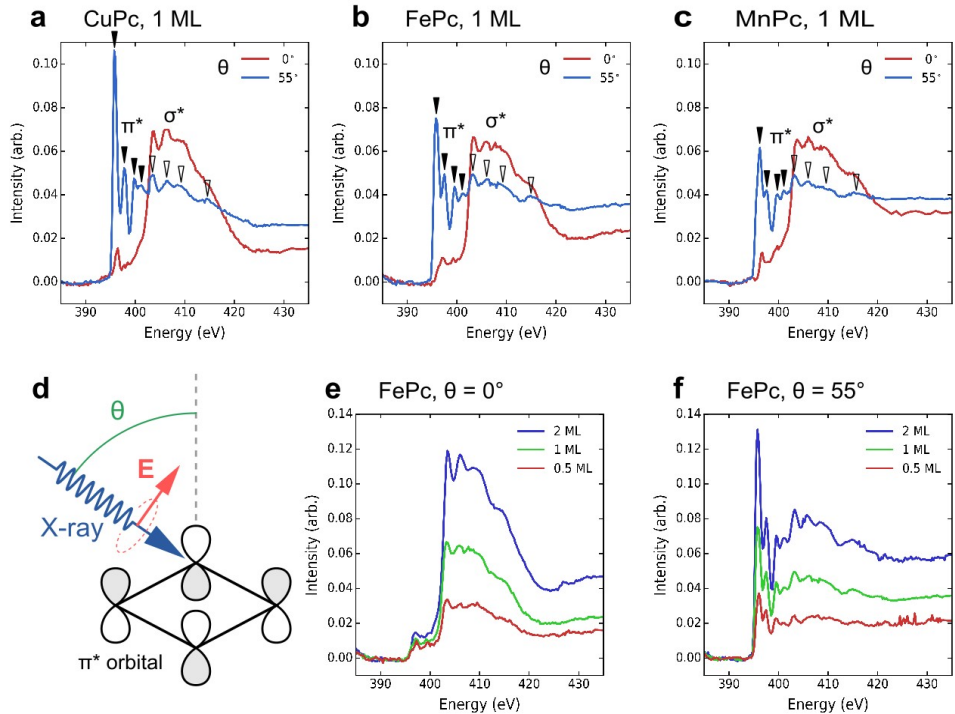

**Figure S4 | a-c**, N $K$-edge XAS spectra of CuPc (**a**), FePc (**b**), MnPc (**c**) monolayers on the (√7×√3)-In surface obtained with $\theta$ = 0° (red) and $\theta$ = 55° (blue). Filled (open) triangles indicate the positions of peaks corresponding to excitations from N 1$s$ to π* (σ*) molecular orbitals. **d,** Schematic illustration for the configuration of the XAS/XMCD measurement of MPc molecules. **e, f,** N $K$-edge XAS spectra of FePc at $\theta$ = 0° (**e**) and $\theta$ = 55° (**f**) measured at different coverages.

**Details of the sum rule analysis**

The effective spin magnetic moment $\langle m_S^{\text{eff}}(\theta) \rangle$ is obtained from the XAS data by using the sum rule [14],

$$\langle m_S^{\text{eff}}(\theta) \rangle = -n_h \frac{9p(\theta) - 6q(\theta)}{r(\theta)} \mu_B, \tag{7}$$

where $n_h$ is the number of holes in the *d* orbitals (assumed to be 1, 4, and 5 for CuPc, FePc, and MnPc, respectively) and $\mu_B$ is the Bohr magneton. The three functions *p(θ)*, *q(θ)*, and *r(θ)* are given by

$$p(\theta) = \int_{L3} [I^+(E,\theta) - I^-(E,\theta)] dE, \tag{8}$$

$$q(\theta) = \int_{L3,2} [I^+(E,\theta) - I^-(E,\theta)] dE, \tag{9}$$

$$r(\theta) = \int_{L3,2} [I^+(E,\theta) + I^-(E,\theta) + I^0(E,\theta)] dE. \tag{10}$$

Here, $I^+(E,\theta)$ and $I^-(E,\theta)$ are the XAS intensities measured with the photon helicity parallel and antiparallel to the applied magnetic field, respectively. $I^0(E,\theta)$ corresponds to the XAS intensity measured with a linear photon polarization parallel to the magnetization, which is approximated as described later. Note that *r(θ)* is an isotropic term and thus is independent of θ in principle.

The quantity $\langle m_S^{\text{eff}}(\theta) \rangle$ is composed of two terms:

$$\langle m_S^{\text{eff}}(\theta) \rangle = 2\langle S \rangle + 7\langle T(\theta) \rangle, \tag{11}$$

where $\langle S \rangle$ is the isotropic spin component and $\langle T(\theta) \rangle$ is the intra-atomic dipolar moment. The latter reflects the anisotropic distribution of spins within an atom and, in a system with a symmetry higher than $D_{2h}$, cancels out at the 'magic angle' of θ = 54.7°[15]. This cancellation of the $\langle T(\theta) \rangle$ term at the magic angle has been demonstrated previously for CuPc and FePc thin films on noble metals[16,17]. For the discussions on the spin magnetic moments in the main text, we used θ = 55° to minimise the contribution from the $\langle T(\theta) \rangle$ term.

The *r(θ)* term was evaluated differently for CuPc and for FePc/MnPc. In the case of CuPc, a hole is present only in the $d_{x^2-y^2}$ orbital. Owing to its $C_4$ symmetry, $I^0(E, 0°)$ is expected to be zero and $I^+(E,\theta) + I^-(E,\theta)$ has a large angular dependence of the form $f(\theta) = (1 + \cos^2\theta)/2$[17]. Hence, *r(θ)* for CuPc is given by

$$r_{\text{CuPc}}(\theta) = \frac{1}{f(\theta)} \int_{L3,2} [I^+(E,\theta) + I^-(E,\theta)] dE \tag{12}$$

In the cases of FePc and MnPc, the angular dependence of $I^+(E,\theta) + I^-(E,\theta)$ is much smaller than that of CuPc, and $I^0$ can be approximated by $I^0 = (I^+ + I^-)/2$. Then, $r(\theta)$ is given by

$$r_{\text{FePc,MnPc}}(\theta) = \frac{3}{2} \int_{L_{3,2}} [I^+(E,\theta) + I^-(E,\theta)]dE. \tag{13}$$

Figure S5a shows the $I^+(E,\theta) + I^-(E,\theta)$ of CuPc on the ($\sqrt{7}\times\sqrt{3}$)-In surface, displaying well-separated $L_{3,2}$ peaks. An additional peak labelled as $A$ is observed around 940 eV in the curve for $\theta = 55°$. This is assigned to the transitions from $2p$ to $4s$ states[18] and disappears at $\theta = 0°$. Since the presence of this peak prevents us from defining an atomic background appropriately, we used $r_{\text{CuPc}}(0°)$ instead of $r_{\text{CuPc}}(55°)$ to calculate $\langle m_S^{\text{eff}}(55°)\rangle$. For FePc and MnPc, such an additional peak was not observed, as shown in Figs. S5b and S5c. For each MPc, an atomic background was simulated by integrating two Voigt functions placed at the $L_{3,2}$ peaks and was subtracted from the XAS spectrum before applying the sum-rule. The results of XAS integrations are also plotted by red curves in Figs. S5a-c where the small circles indicate the upper bound of the $r(\theta)$ integration range.

Figures S5d-S5f show the XMCD spectra, $I^+(E,55°) - I^-(E,55°)$, and their integrals evaluated for the three MPcs. The integration ranges for $p(\theta)$ and $q(\theta)$ are indicated by the small squares and triangles, respectively. In the case of MnPc (Fig. S5f), XMCD signals from $L_{3,2}$ peaks are partially overlapped and the upper bound of the integration for $p(\theta)$ is not obvious. For practical solution, we used the energy at the local minima (labelled as $B$ at 645.5 eV) between the $L_{3,2}$ peaks in the XAS

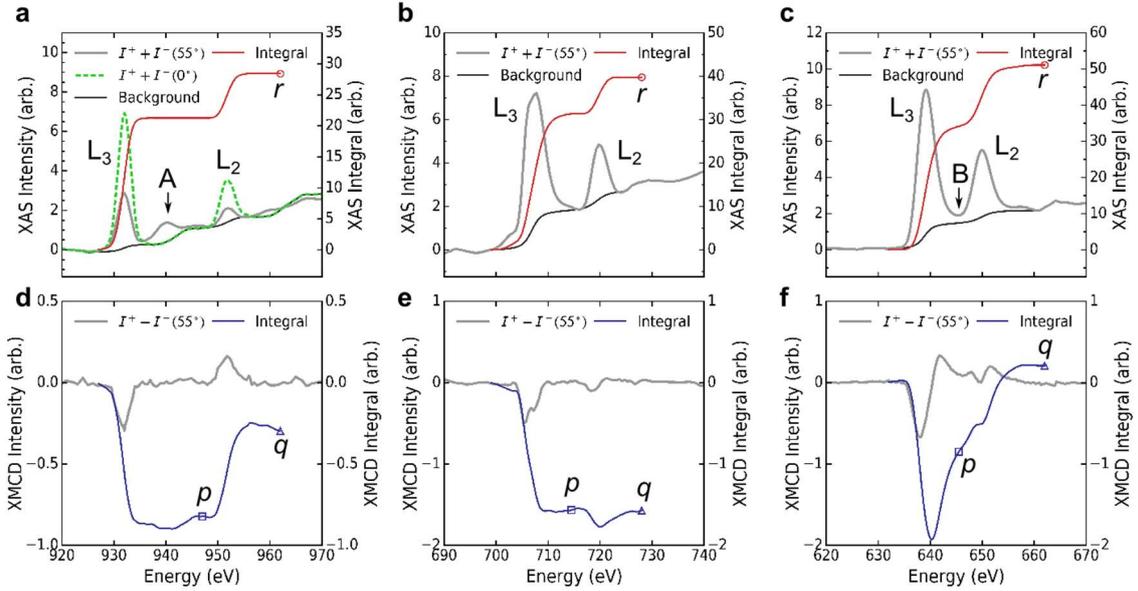

**Figure S5** | **a-c**, XAS spectra ($I^+ + I^-$) of CuPc (**a**), FePc (**b**) and MnPc (**c**) at $\theta = 55°$ (gray curves). The atomic backgrounds are plotted in thin black curves. The integrals of XAS spectra are plotted in red and the points defining the $r(\theta)$ integration is marked by circles. For CuPc, the XAS integral was calculated from the spectrum for $\theta = 0°$ plotted with a green dashed curve in the panel **a**. The small peak labeled as $A$ in **a** is related to transitions from $2p$ to $4s$ states. In the panel **c**, the arrow $B$ indicates the local minimum of the XAS spectrum used to determine the upper bound of the integration for $p(\theta)$. **d-f**, XMCD spectra ($I^+ - I^-$) of the three MPCs. The points defining the integrations for $p(\theta)$ and $q(\theta)$ are marked by squares and triangles, respectively.

curve (Fig. S5c). If we move this upper bound by ±1 eV, the resulting $\langle m_S^{\text{eff}} \rangle$ value changes by $\mp 10\%$. A theoretical study shows that $\langle m_S^{\text{eff}} \rangle$ obtained from a sum-rule analysis on Mn$^{2+}$ is underestimated by 32% [19].